# CINNAMONS: A COMPUTATION MODEL UNDERLYING CONTROL NETWORK PROGRAMMING

Kostadin Kratchanov

Department of Software Engineering, Yaşar University, Izmir, Turkey
kostadin.kratchanov@yasar.edu.tr

## ABSTRACT

*We give the easily recognizable name "cinnamon" and "cinnamon programming" to a new computation model intended to form a theoretical foundation for Control Network Programming (CNP). CNP has established itself as a programming paradigm combining declarative and imperative features, built-in search engine, powerful tools for search control that allow easy, intuitive, visual development of heuristic, nondeterministic, and randomized solutions. We define rigorously the syntax and semantics of the new model of computation, at the same time trying to keep clear the intuition behind and to include enough examples. The purposely simplified theoretical model is then compared to both while-programs (thus demonstrating its Turing completeness), and the "real" CNP. Finally, future research possibilities are mentioned that would eventually extend the cinnamon programming into the directions of nondeterminism, randomness and fuzziness.*

## 1. INTRODUCTION

This paper introduces a new computation model called **Core Control Network Programming**, or **Core CNP**. We introduce it to serve as a theoretical basis for Control Network Programming [1-4]. Actually, to make the new computation model easier to distinguish (and influenced by the popularity of Pokémon), we will slightly twist the name "Core CNP" into "**Cinnamon programming**". As it is intended for formal mathematical study, it is a strictly minimal version of CNP - it has the same structure, but with only the most basic and fundamental features and primitives necessary. Sections 2 and 3 are devoted to the syntax and semantics of the new computation model, respectively. In Section 4 we show the Turing-completeness of cinnamon programming. Section 5 includes a concise description of more advanced features that distinguish the "real" CNP from the core CNP while Section 6 sketches some ideas for future study.

In addition to developing theoretical foundations for CNP, an equally important aim of the presentation is to make the syntax and semantics of Control Network (CN) programs unambiguous and clear for the programmers employing CNP.

### 1.1. Background and Perspective

As Jones explicitly maintains in his Preface to [5], the two major computer science fields – Theory of Computation (which splits into Computability and Computational complexity), and Programming languages (including Syntax and Semantics) – have much to offer each other. This overlap of concepts, approaches, and results can be described as a trend observed in the recent decades, some examples being [6-8]. We would like to follow this trend here and use the terminology from both areas.

The purpose of a computation is to solve a **problem**. Clearly, we must first make precise what is meant by a problem (also called **computational task**). Currently, there are four well-established main types of computational tasks considered in the computer literature: decision problems, function problems, search problems, and optimization problems.

There are numerous approaches to the definition of **computation** and, respectively, to answering the question what is computable. Examples of some of the most famous approaches are: recursive functions (originating from the notion of functions in mathematics and based on the intuitive idea of what operations can be reasonably considered to be easily computable), lambda calculus (underlying functional programming), Turing machines (a maximally simplified version of a computer), register machines and RAM machines (underlying machine languages and assembly languages), GOTO-programs and WHILE-programs (theoretical foundation for higher-level imperative programming languages), first-order logic (underlying logic programming). A really striking result and one of the most important achievements of computer science is the fact that all these very different approaches ultimately lead to the same classification of decision problems, respectively function problems, with respect to their computability. Computability (solvability) turns out to be a natural intrinsic property of problems independent of the formalism used in the corresponding approach. These formalisms are referred to as models of computation. A **computation model** is called Turing-complete if it is equivalent in its computational power to a Turing machine (and, correspondingly, to any other of these universal models).

Under each of these approaches, a model of computation is defined as a formal mathematical object, followed by the concept of computation for that model and the notion of what a given specific model computes. Assume that Turing machines are our computation model. Turing machines – acceptors are used for solving decision problems, while Turing machines – transducers are used for solving function problems. A Turing machine – acceptor solves a decision problem by accepting or rejecting the element. A particular Turing machine – transducer calculates a partial function by producing an output element for an input element.

Note that, ultimately, every Turing machine (including its program), or WHILE-program, etc. has a unique description in some properly defined language. Therefore, we can talk about the language of Turing machines, the language of WHILE-programs, and so on. A particular Turing machine is an element (a well-formed sentence) of the language of Turing machines.

The description of a language is split into **syntax** (form) and **semantics** (meaning). Semantics reveals the meaning of syntactically valid strings in a language. There is a wide range of semantics proposed for programming languages (see, e.g., [4]). Two of the major approaches are outlined next. **Operational semantics** describes how a computation is performed internally in the corresponding computation model (often also called an abstract machine) – that is, how the program is interpreted (executed) in the computation model. In contrast, in **denotational semantics** we give meaning to the program by specifying the external behavior achieved by the computation, e.g., the partial function computed.

The above means that introducing a formal model of computation as a mathematical object (e.g., Turing machines) and defining how the computation in its terminology is performed (how a Turing machine operates and what it computes) is actually equivalent to introducing a language (the language of Turing machines) and specifying its syntax and semantics.

### 1.2. Cinnamons

This paper introduces a new computation model called **Core Control Network Programming**, or **Core CNP**.

What in traditional programming corresponds to a program, is called here a **cinnamon** (the formal definition of a cinnamon is given in the next section). A traditional program is a string in the corresponding programming language. In contrast, a cinnamon is not simply a string but rather a 'control network' – a set of graphs. Cinnamon programming is a type of graphical (visual) programming. We believe graphical programming is clearer and more natural for human programmers [1,4], and this results in a faster process of developing a solution to the problem in hand. Of course, at a lower level the CN is transformed into a string in an appropriate language which is manipulated by the CNP compiler. However, in this presentation we prefer to stay at the higher and more abstract level.

It is customary in mathematics to discuss objects without specifying their representation (talking about sets of objects, functions between sets of objects, etc.). In computation theory the representation of objects often plays a central role [10]. However, for clarity and simplicity, we will still define, as much as we can, the notions and concepts we need using sets without regard for the representation of their elements. Actually, computational tasks refer to objects that are represented in some canonical way. The two such main and best studied alternative representations are strings of symbols and natural numbers. In our presentation, when that must be made explicit we have chosen to work with natural numbers, and correspondingly functions on natural numbers.

## 2. SYNTAX

The syntax of the "language" of cinnamons is defined below.

### 1.2. Definitions

Let $SV = \{FINISH, RETURN\}$ be a set of two distinct elements called **system vertices (system nodes, system states)**, $FINISH \neq RETURN$. A **graph** is an ordered 6-tuple $G = <S, A, source, target, L, label>$ consisting of: a nonempty finite set, $S$ of elements called **ordinary states (ordinary nodes, ordinary vertices)**, $S \cap SV = \emptyset$; a finite set, $A$ of **arrows**; two functions *source: $A \rightarrow S$* and *target: $A \rightarrow (S \cup SV)$* mapping an arrow $a \in A$ into its **source** $s \in S$ and **target** $s' \in (S \cup SV)$ respectively; a set, $L$ of **labels**; and a function *label: $A \rightarrow L \cup \{\lambda\}$* assigning a label $label(a) \in L \cup \{\lambda\}$ to each arrow $a \in A$. A **state** is either an ordinary state, or a system state. For a given $s \in S$, we will denote by *out(s)* the set of all arrows with source s (called also arrows outgoing from s). An **initialized graph** is an ordered pair $<G, s_o>$ where $G$ is a graph and $s_o$ is a selected state, $s_o \in S$ called the **initial state**. Note that *FINISH* and *RETURN* have no outgoing arrows, and that these system states cannot be initial. We will call a graph, $G$ an **ordered graph** if the set *out(s)* for any given vertex $s \in S$ is linearly ordered. An ordered graph which is initialized, is called an **ordered initialized graph**.

We will represent initialized graphs graphically, using the following graphical symbols:

| Graphical symbol | 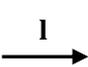 | 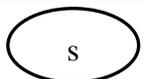 | 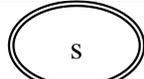 | 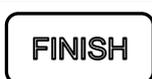 | 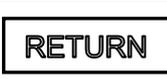 |
|---|---|---|---|---|---|
| Element | Arrow with label l | (Ordinary) vertex | Initial vertex | *FINISH* vertex | *RETURN* vertex |

In the *SpiderCNP* IDE, the graphical symbol for an initial node is an oval in green: 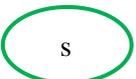 .

*SpiderCNP* [4,11,12] is an integrated development environment for developing and running CN programs. In particular, CNs can be created, tested and edited using a built-in graphic editor. All illustrations of CNs in this paper are screenshots from this IDE. *SpiderCNP* was developed by T. Golemanov.

A **subnet** is an ordered pair $<G,P>$ where $G$ is an ordered initialized graph, and $P$ is a list (possibly empty) with elements called **formal subnet parameters**.

Let *Vars* be a countable set. Its elements will be called **variables**. We will denote the elements of *Var* as $x_0, x_1, x_2, ...$ Of course, *Vars* could be the set N of natural numbers itself.

A **cinnamon** (a substitution for Simple CN) is comprised of a nonempty finite set of subnets one of which is identified as the **main subnet**. Formally, a cinnamon, $\Sigma$ is an ordered pair $<N, m>$ where $N$ is a set of subnets satisfying conditions 1) - 2) below, and $m \in N$ is its main subnet.

This set of subnets must satisfy the following two properties:

1) The sets of vertices of the (graphs of the) subnets are mutually nonintersecting.
2) All subnets share the same label set, $L$. This label set consists of two types of elements: **primitives**, and **invocations** (the latter also called **subnet calls**) described below.

Each primitive in the cinnamon may be from one of the following four types:

    a)    *clear(x)*    (also denoted *x:=0*)
    b)    *copy(x,y)*    (also denoted *y:=x*)
    c)    *inc(x)*    (also denoted *x:=x+1*)
    d)    *if nonEq(x,y)*    (also denoted *if x≠y*)

Primitives *clear*, *inc*, and *copy* are called **elementary action primitives**. Primitive *if nonEq* is called the **elementary test primitive**.

Above, $x$ and $y$ are variables. Note that $\Sigma$ consists of a finite set of subnets, each of which has a finite set of labeled arrows. Therefore, the number of variables used in it is also finite. We will denote by $Vars^\Sigma$ the set of all variables used in the primitives of the cinnamon $\Sigma$. If $\eta$ is a subnet of a cinnamon $\Sigma$, than $Vars^\eta$ denotes the set of variables of $\eta$. The sets of variables of two distinct subnets are nonintersecting. Clearly, $Vars^\Sigma$ is the union of $Vars^\eta$ for all subnets $\eta$ of $\Sigma$, $Vars^\Sigma$ is finite, and $Vars^\Sigma \subset Vars$.

An invocation has the form:

3) *CALL$\eta$ ($a_0, a_1, ..., a_{n-1}$)* where $\eta$ is a subnet of the cinnamon $\Sigma$, and $a_0 - a_{n-1}$ are called **actual subnet parameters**. Each $a_i$ is either a variable from $Vars^\Sigma$, or a constant (a natural number).

The list of actual subnet parameters may be empty, in which case we simply write *CALL$\eta$*. The number of actual parameters in the subnet call equals the number of formal parameters in the subnet definition.

Both direct and indirect recursion are allowed. For example, it is possible that the invocation *CALL($\eta$)* is located in subnet $\eta$, that is, subnet $\eta$ is called from itself.

According to our definition above each arrow has a single label which is an elementary primitive or a subnet call. This assumption is imposed for simplicity. In practice, it is more convenient to allow an arrow's label to be a finite, possibly empty, string of primitives and/or invocations. If using the new notation, 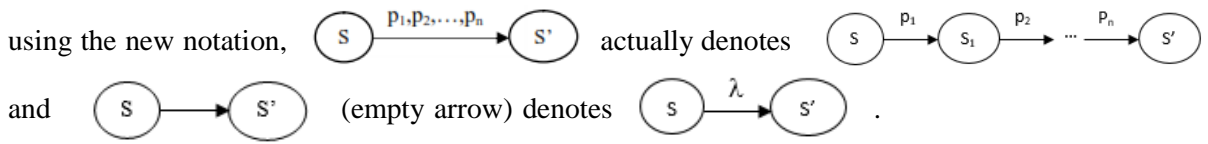 .

An example of a cinnamon is shown in Figure 1.

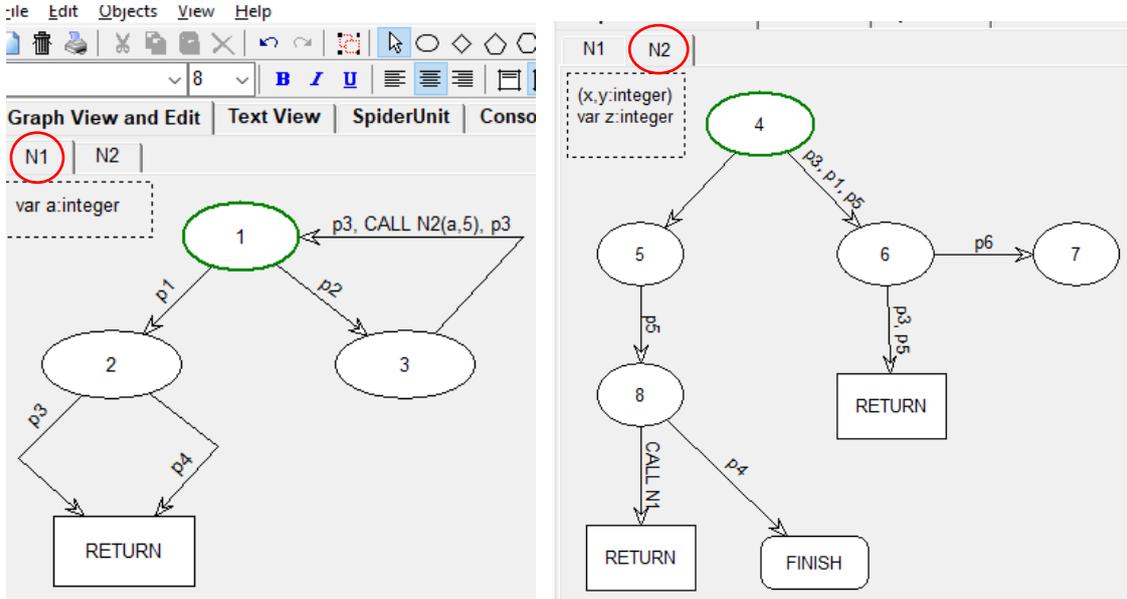

Figure 1 Cinnamon example 1

Here, the cinnamon consists of two subnets, *N1* and *N2*. N1 is the main subnet. It includes the ordinary states {*1, 2, 3*}, as well as a *RETURN* system state. The initial state is *1*. Primitives used in *N1* are {*p1, p2, p3, p4*}; the invocation *CALL N2(a,5)* is also used. In the invocation, *a* is an actual subnet parameter; the second parameter in this call is the constant 5. Subnet *N2* uses the primitives {*p1, p3, p4, p5, p6*}, and the invocation *CALL N1*. Its set of states is {*4, 5, 6, 7. 8*}. *N2* also includes two *RETUN* states and one *FINISH* state. Each one of the primitives *p1 – p6* must be of the forms discussed above, i.e., either an elementary action primitive or elementary test primitive. Subnet *N1* has no formal subnet parameters while subnet *N2* has two formal subnet parameters, *x* and *y*. This is an example of indirect recursion: subnet *N2* is called from within subnet *N1*, and *N1* is called from within *N2*. Clearly, invocating the main subnet form the same or other subnet is not a hindrance.

As emphasized, the arrows going out of a given node, are linearly ordered. For convenience we accept that on the graphical representations the order of the arrows is from the left to the right and up – down. For example, the arrow *1 → 2* in *N1* is before arrow *1 → 3*. If using this default rule on the graphical representation is difficult, we may also show the order explicitly, as in Fig. 2. Here, arrow *1 →3* precedes arrow *1 → 2*.

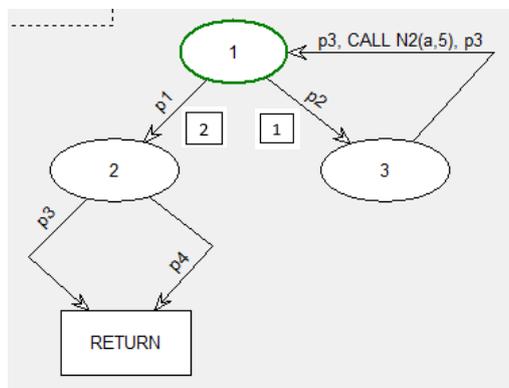

Figure 2 Indicating order of arrows

In our definition of a cinnamon above we have chosen a minimal set of allowed elementary primitives consisting of the specified three elementary action primitives (*x:=0, y:=x*, and *x:=x+1*) and one test primitive (*if x≠y*). As we will see further, this minimal set is functionally complete in the sense that these primitives are enough for implementing any computation. It would have been possible to choose alternative sets, e.g., action primitives {*x:=0,* y:=*x+1,* y:=*x÷1*}. As far as the test primitive is concerned, its condition could have been replaced by any one of conditions =, <, >, ≤, ≥ (similarly to [6,13] in the case of while-programs). For our considerations, however, our choice of elementary primitives proves to be the most convenient.

It is also possible to choose a different set of elementary primitives if working with a based-on-strings representation of data. Let *A* be a fixed alphabet with at least two different symbols. We could, for example, choose the following set of elementary primitives: *clear(l)* (creating an empty string *l=λ*), *sep(l,h,t)* (separating a string *l* into its head *h* and tail *t*), *cons(h,t,l)* (constructing a new string *l* with head *h* and tail *t* = adding a new symbol to the beginning of a string), *if Eq(a,a1)* (comparing symbols), *if Empty(l)* (*if l = λ*).

It is worth mentioning that the concept of a cinnamon is an elaborated, filled with flesh version of an introduced much earlier but not so widely used skeletal notion – the notion of a recursive automaton / recursive finite-state automaton / recursive transition network / recursive control graph [14-23]. It is a computation model equivalent in computational power to nondeterministic push-down automata. Basically, the difference between an automaton and a recursive automaton is that the latter is a set of nondeterministic automata in which transitions are labeled either by an element of the input alphabet (a primitive in the case of cinnamons) or by the name of a nondeterministic automaton from the set (a subnet in the case of cinnamons). Cinnamons are a more complicated concept than recursive automata – for example, primitives and subnets have parameters, the set of outgoing arrows is ordered, the behavior in the basic definition is deterministic, there are concepts like failure and backward execution (the latter differences are of semantic nature).

As is customary in computation theory literature, we can allow for convenience the usage of macro primitives. A **macro definition** is a sequence of primitives. For example, the sequence <*clear(z), nonEq(x,z)*> can be considered as a macro definition for the macro statement *nonZero(x)* which would be a new test primitive. Such an abbreviation will be called a **macro statement** (plural: **macros**).

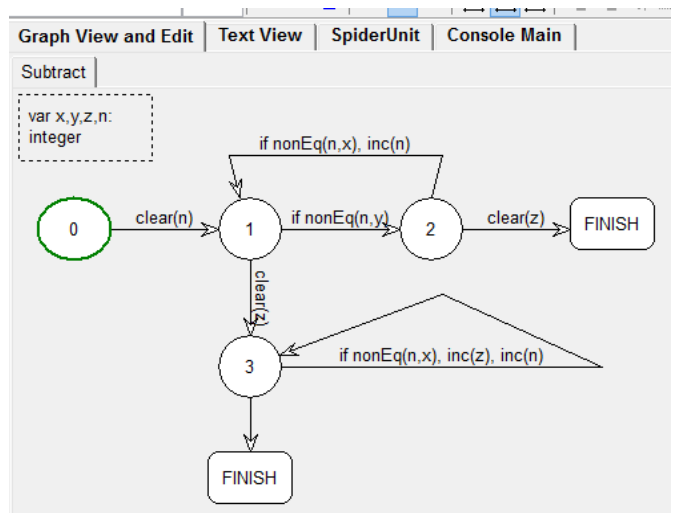

Figure 3 Cinnamon for subtraction

## 2.2 Examples

A second example of a cinnamon is shown in Fig. 3. It consists of a single (main) subnet *Subtract*. After understanding the semantics (computation) of cinnamons, one can convince himself that this cinnamon computes the function subtraction of natural numbers; more precisely,

$$z = \begin{cases} x - y, if\ x \geq y; \\ 0, if\ x < y. \end{cases}$$

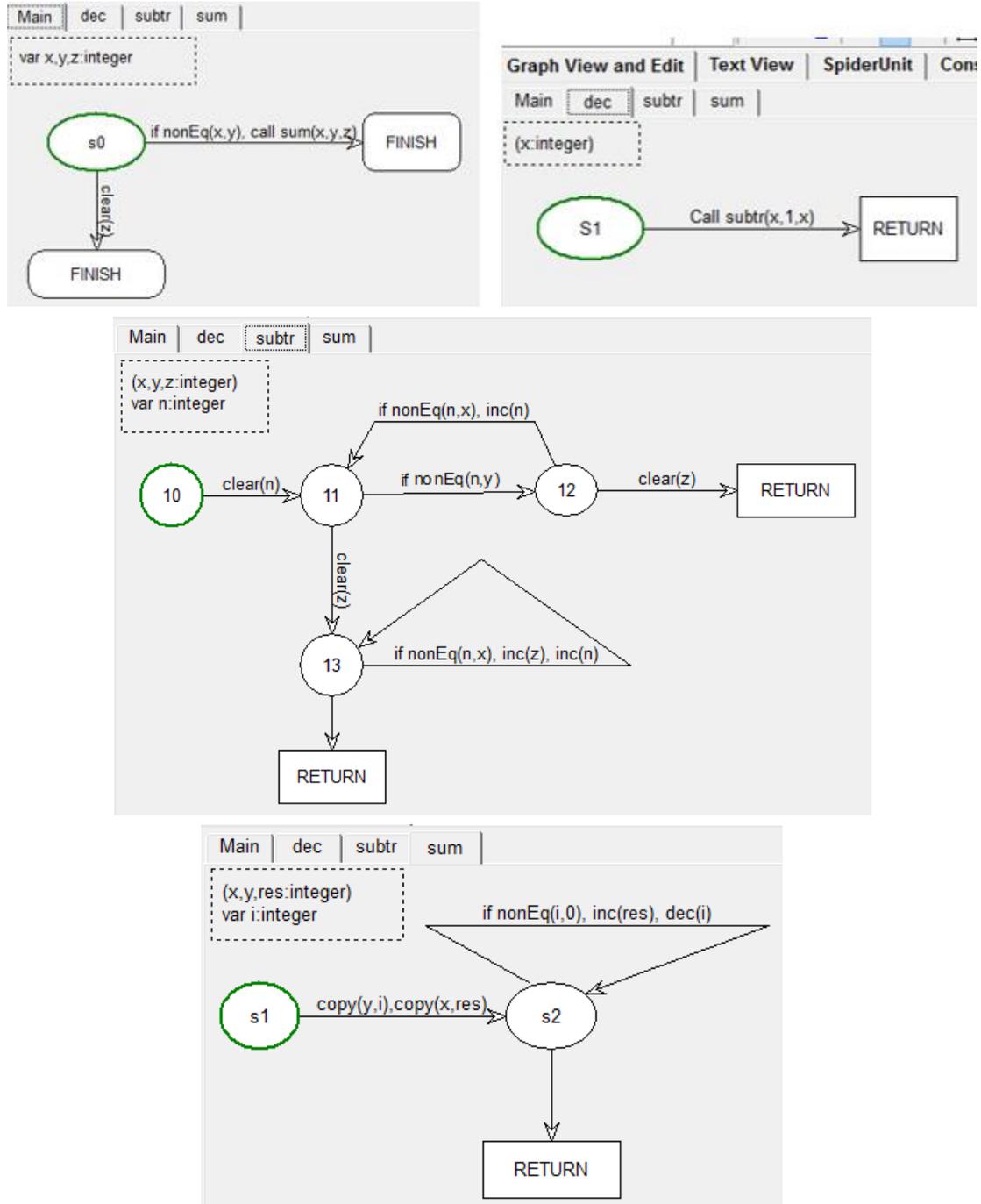

Figure 4 A larger example

Macros are one method for separating parts of a cinnamon into a named module of some sort and re-use them. Another, more powerful method is using subnets. In the following example (Figure 4), we use a modified version of the previous example – the cinnamon for subtraction, but now it is not a complete cinnamon but rather a subnet within a cinnamon. The cinnamon has four subnets. One is subnet *subtr* already discussed. Subnet *dec* 'defines' a new 'operation' – decrement, using the subnet for subtraction. Using *dec*, subnet *sum* 'defines' another new 'operation' – summation. The overall control structure is determined by the main subnet. This example actually implements the function

$$if\ x \neq y\ then\ z := x+y\ else\ z: = 0$$

Note that ordering of outgoing arrows from a certain node plays an important role. For example, according to the default ordering of arrows outgoing from node *s0* in subnet *main* arrow *s0*→*FINISH* labeled *if nonEq(x,y), call sum(x,y,z)* precedes the arrow labeled *clear(z),* and the loop *s2*→*s2* in subnet *sum* precedes arrow *s2*→*RETURN*.

Our next example (Fig. 5) illustrates a cinnamon based on the definition with data representation using strings instead of natural numbers. Here, *copy(t,l)* is a macro statement with macro definition <*sep(l,h,t, cons(h,t,l)*> (*l := t*).

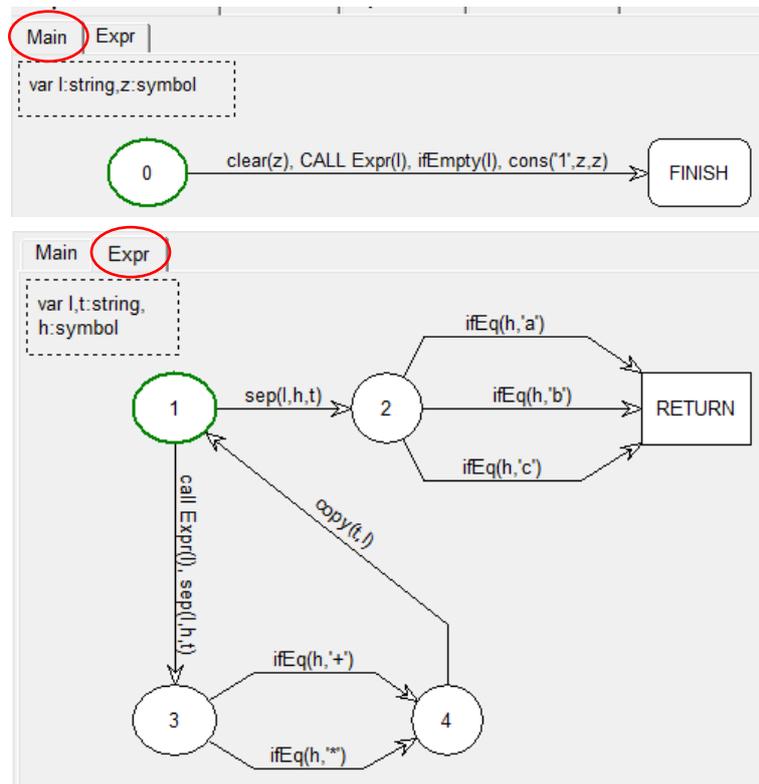

Figure 5 Recognizer of arithmetic expressions

Assuming that the alphabet *A* includes symbols *a, b, c, +,* and *\**, this cinnamon plays the role of a recognizer for arithmetic expressions defined by the (ambiguous) grammar

$$E \rightarrow E + E\ /\ E * E\ /\ a\ /\ b\ /\ c$$

Equivalently, this grammar can be defined by the syntax diagram in Fig. 6. The structure of the graph of the cinnamon duplicates that of the syntax diagram. The left recursion has been avoided.

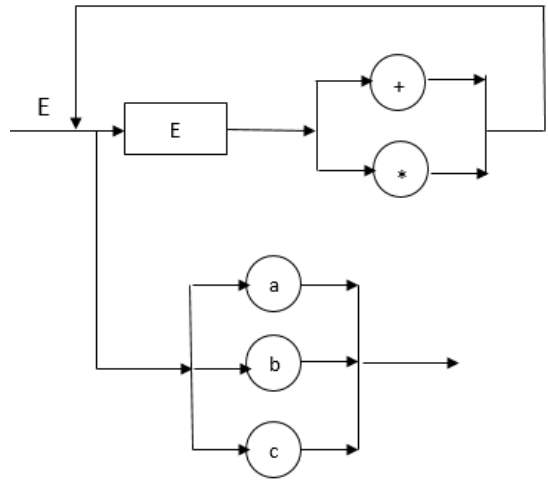

Figure 6 Syntax diagram of the grammar of arithmetic expressions

Note that the list of elementary primitives in the definition of a cinnamon includes no 'control' primitives. This contrasts with the other well-known computation models such as while-programs, recursive functions, etc. where one can find 'control' components such as while-loops, conditional statements, functions of primitive recursion and unbounded minimization, etc. The primitives in our definition are elementary tests or actions, and the control is embodied in the structure of the subnets. In the 'real' CNP, however, primitives are user-defined and can be arbitrary procedures defined in the underlying language – they can include loops, etc. In the triple general computational model terminology [22], variables and primitives define the operational unit, while the CN defines the control unit. As emphasized in [4], "Primitives + Control Network = Control Network Program". In cinnamon programming primitives are built-in. In real CNP primitives must be defined and therefore a CNP project in *SpiderCNP* or the *Bouquet* cloud development environment [4] involves two major files – *SpiderUnit* and *SpiderNet*; in cinnamon programming *SpiderUnit* is not needed.

Above, a cinnamon was defined abstractly using purely mathematical notions (such as sets and graphs) and notations. Alternatively, we could've had defined a formal language, or a programming language in which a cinnamon is specified. For our purposes in this paper the more abstract approach is preferable as it presents the concepts in a much more clear and neat manner. In reality, for developing and running CN programs one usually uses an integrated development environment such as *SpiderCNP* or *Bouquet* [4]. In *SpiderUnit*, two equivalent representations of the CN exist – a graphical representation and a corresponding textual representation. Typically, the 'programmer' creates and modifies the CN using its graphical representation manipulated by the built-in graphical editor, while the textual representation of the same CN is used in the files with which the CNP compiler and other software modules work.

## 3. SEMANTICS

In the previous section we presented the syntax of the 'language' of cinnamon programming, in other words, what syntactically a cinnamon is. As already discussed this is technically not a language as a cinnamon, unlike a program, is not a string but rather a set of graphs.

Now we turn our attention to the semantics of a cinnamon – in other words, what a cinnamon 'computes'. This specification of the computation process can be considered as an operational semantics. One can also consider a partial function which is defined by this computation thus saying that a cinnamon as a syntactic construct specifies this function – and that can be considered as a denotational semantics.

## 3.1 Informally on the computation in a cinnamon

We will describe first the 'computation' performed in a cinnamon, and then will address the semantics of the model more rigorously. Our exemplary cinnamon (which is a simplified version of the 'technical example' from [22]) is shown below.

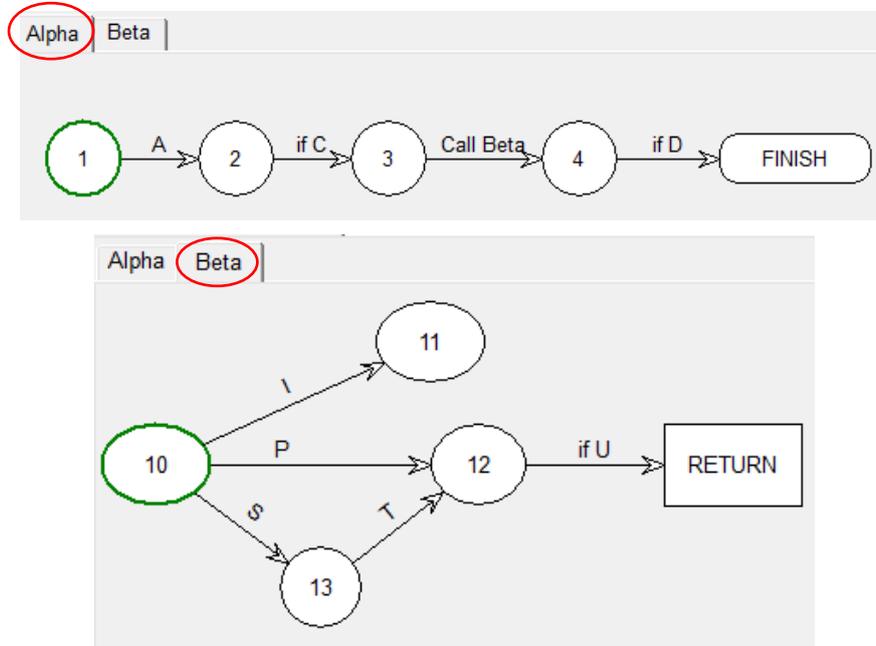

Figure 7 Technical example

The computation starts from the initial state of the main subnet by 'forward' traversal along the arrows, and executing each primitive on the way. An elementary test primitive can be executed successfully, or unsuccessfully. After unsuccessful execution, the direction of the traversal is switched to 'backwards'. How a primitive is executed forwards and backwards is given in its definition. The traversal strategy is an extended version of backtracking. When the control is returned to a state, the next not attempted yet outgoing arrow is tried. If no non-attempted outgoing arrows exist, the control goes backwards through the arrow along which the state was reached. A very interesting point of cinnamon programming is that the backward traversal can enter backwards a subnet, and continue backwards along the arrow used before. This is not possible in traditional programming because when completing a procedure all information from inside the procedure is lost. In contrast to traditional procedure/function call in programming languages, in cinnamon programming the data are not restored from a data stack but are restored by backwards execution. If the traversal reaches a *FINISH* node than the computation finishes successfully. If the traversal gets stuck in the initial node of the main subnet then the computation finishes unsuccessfully – no solution has been found.

One possible traversal is shown in Fig. 8. The execution starts from the initial state, *1* of the main subnet, *Alpha*. Action primitive A is executed forwards, then test primitive *if C* is executed successfully. Next invocation *Call Beta* is executed, after which the control is in the initial state, *10* of subnet *Beta*. Primitive *I* is now executed, and control moves to state *11*. However, state 11 has no outgoing arrows, and mode of traversal changes to 'Backwards'. Primitive *I* is now executed backwards and control is back in state *10*. The next in order not-attempted outgoing arrow is the one with primitive *P* which is now executed. Then test primitive *if U* is successfully executed, *RETURN* state is reached, and control jumps back to subnet *Alpha*, more specifically to state *4*. Test primitive *if D* follows, but its execution is unsuccessful. Therefore, backwards execution is triggered. Primitive *if D* is executed backwards, and control returns to state *4*. There are no other outgoing arrows from *4*. Therefore, subnet *Beta* is entered backwards. The system remembers that the last executed arrow before returning from subnet *Beta* was the one with label *if U*. Now, if U is executed backwards. State *12* has no remaining not-attempted arrows, so control continues moving backwards and primitive *P* is executed backwards. Now the control is in state *10*. Outgoing arrow *S* has not been attempted, so the mode changes back to 'Forward' and primitive *S*, and then *T* and *if U* are executed forwards. Now, return from subnet Beta is performed and control reaches state *4*. Assume that test primitive *if D* is now successful. Thus, the control reaches state *FINISH,* and the computation completes successfully.

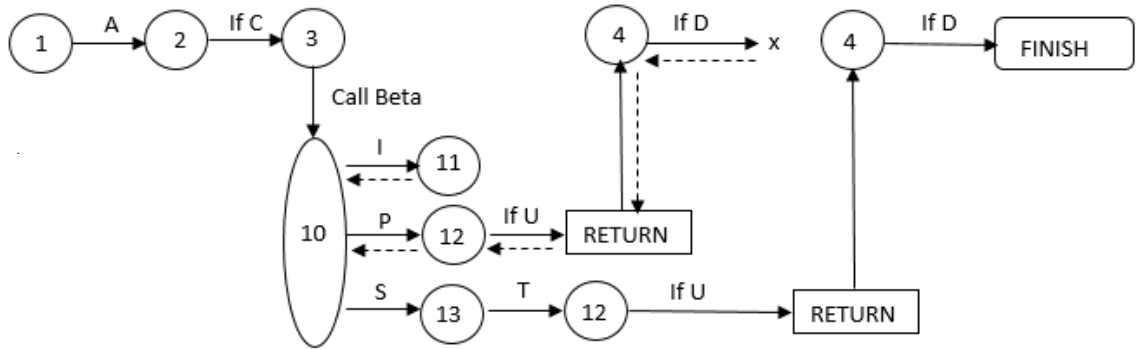

Figure 8 One possible execution in the technical example

Now we can proceed to a formal definition of the cinnamon semantics.

## 3.2 Formal semantics

Let *Vars* be the countable set of variables defined earlier in the Section on Syntax. Let $\Sigma$ be a given cinnamon, and $Vars^\Sigma \subset Vars$ be its finite set of variables. As a matter of fact, we never use values of Vars outside of $Vars^\Sigma$ but the definitions remain simpler for *Var*. We will also need two special '*system variables*' FORW and FAILURE. We assume that $SV = \{FORW, FAILURE\}$, $SV \cap Vars = \emptyset$, and $Varss = SV \cup Vars$.

An **environment,** $\varepsilon$ is a partial function $Varss \rightarrow N$. Terms with similar meaning used in literature are state of computation, store, single-assignment store. If for $v \in Varss$ the partial function σ is defined then the natural number $\varepsilon(v)$ is called the **value** of variable *v*. If $\varepsilon(v)$ is not defined, we will also say that the value of variable *v* is not defined. The values $\varepsilon(FORW)$ and $\varepsilon(FAILURE)$ can only be 0 or 1, therefore the two system variables will more often be called **system flags**. The set of all environments is denoted *Env*.

Informally, $Vars^\Sigma$ imitates the data store (operational unit) in our model. We also need a control unit which in cinnamon programming is much more complex than in most other computation models as cinnamon programming is intended for declarative solution of problems of nondeterministic nature. Following our general strategy in this paper, we will employ mathematical definitions rather than explicitly specifying the abstract machine or use programming style terminology. However, in order to simplify the presentation, we will still use popular notions such as a stack of a particular type of elements for example, instead of describing it in formal mathematical language as a function $N \rightarrow Elements$.

We will define the operational semantics of cinnamons by presenting formally an interpreter. The algorithm of this interpreter is specified by its UML activity diagram shown in Figure 9.

The corresponding virtual machine includes as components two stacks and a number of global variables. Stack *arr_ST* stores the current path of arrows and its elements are ordered pairs of the form <*ARR, Remaining*>. *ARR* can take as values the name of any of the arrows in the cinnamon, as well as the special value *separator* which signals that jump to a subnet has been performed. Stack *subnet_ST* keeps information about the nested subnets invoked. Its elements have the format <state after the subnet call, last value of *Remaining* when the call was made, values of parameters of the calling subnet, values of local variables of the calling subnet>. The global variables are: the system flags *FORW* and *FAILURE*, the current state *STATE* of the CN, the list (stack) *Remaining* of the outgoing arrows for the current state that have not been attempted yet, current arrow *ARR*, list *PARAM* of values of the parameters of the current subnet, list *LOCALS* of the values of the local variables of the current subnet, current primitive *PRIM* to be executed. The following notations have been also used in the activity diagram: *init(SN)* is the initial node of subnet *SN*, *out(STATE)* is the list of arrows outgoing from *STATE*, and *target(ARR)* is the state which is the target of arrow *ARR*.

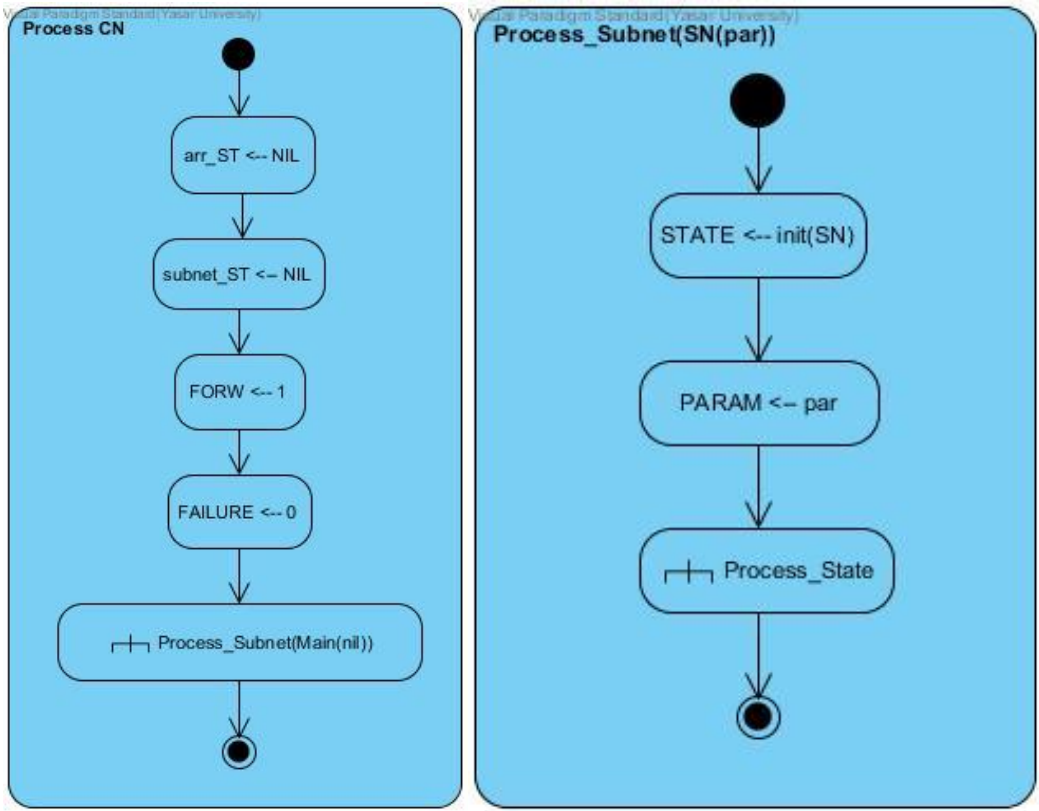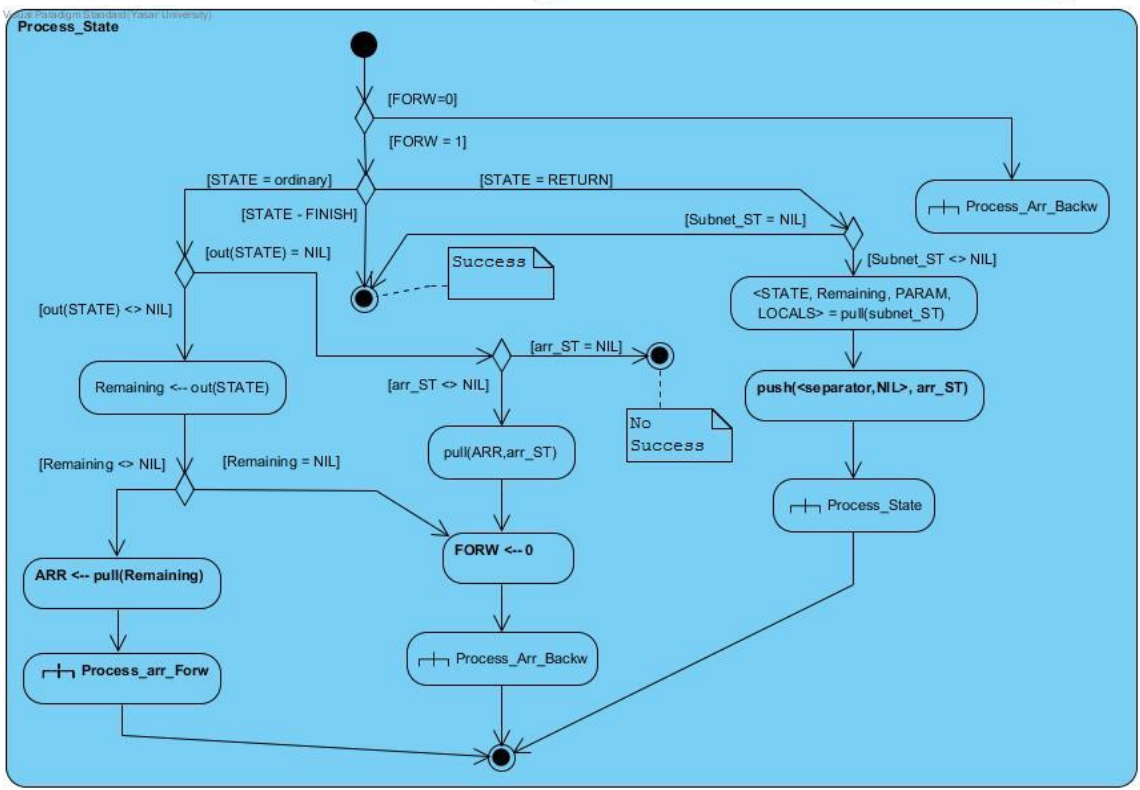

Figure 9a Activity diagram of the interpreter I

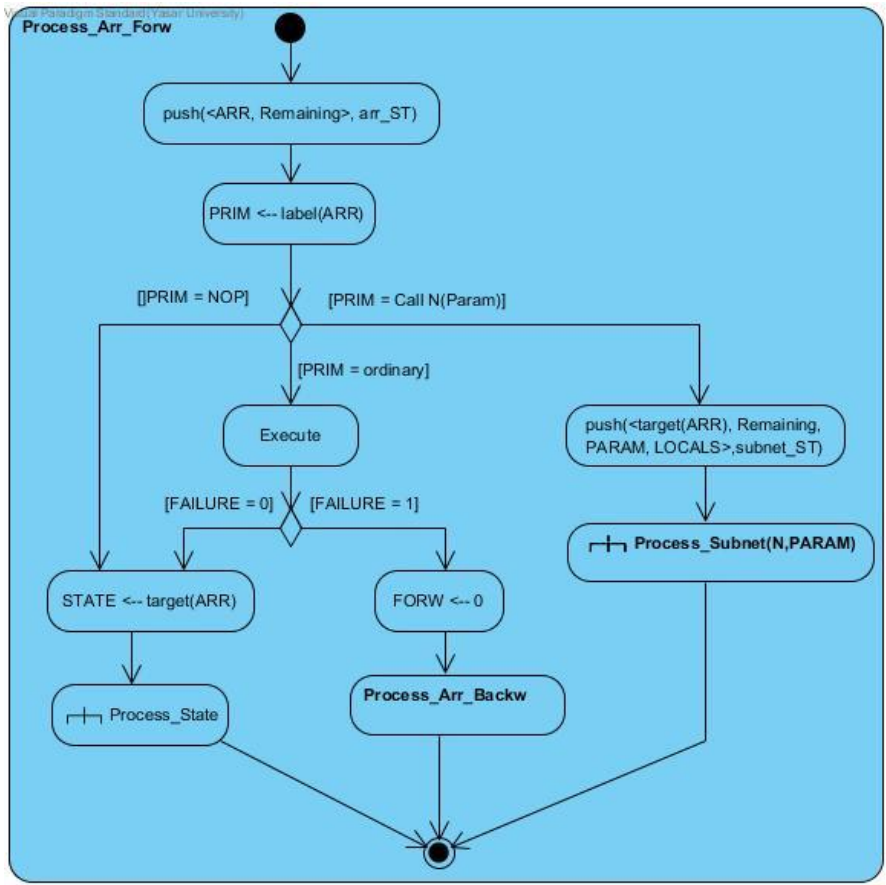

Figure 9b Activity diagram of the interpreter II

The interpreter presented above specifies the operational semantics of the language of cinnamons, and is a theoretical model only. Instead of an interpreter, the *SpiderCNP* IDE contains a recursive-decent-type compiler [4,11,12,22].

Note that the activity diagrams *Process_Arr_Forw* and *Process_Arr_Backw* contain actions called '*execute*'. This stands for 'execute current primitive *PRIM*'. In the former activity diagram the execution is forwards, while in the latter it is performed backwards.

The backwards execution involves restoring the data. That means that normally the action primitives must have two types of action – forward action (when moving forwards) and backward action (when moving backwards and the data are being restored). Recall that the primitives can be of two types: elementary action primitives and elementary test primitives. Only execution of action primitives (*clear*, *copy*, and *inc*) changes the environment. Execution of test primitives does not affect the environment. We can easily define a backward action for *inc*. However, there is no natural way to define a backward action to *clear* and *copy*. Technically, their backward action is empty; in reality, a programmer never uses these two action primitives in positions in the CN where a backward action might be necessary. Values of system flags may be changed by an action primitive, or directly by the control unit.

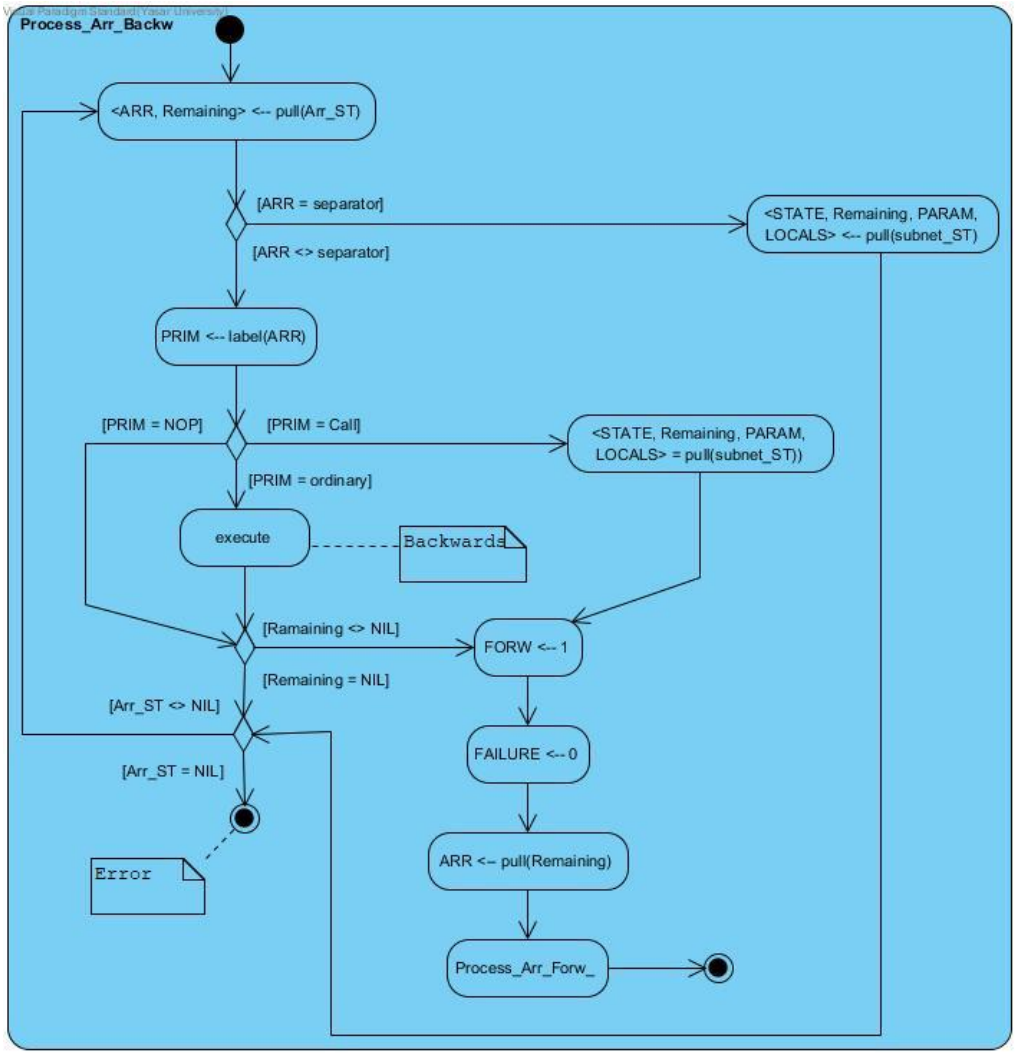

Figure 9c Activity diagram of the interpreter III

If a cinnamon is started in an initial environment $\varepsilon$, then in the course of the 'computation' the control moves in accordance with the activity diagram, and from time to time action primitives will be executed thus changing the values of variables. If and when the cinnamon halts, the final environment will in general be different from $\varepsilon$. We thus interpret a cinnamon $\Sigma$ as a partial function $[[\Sigma]] : Env \rightarrow Env$. The value $[[\Sigma]] (\varepsilon)$ is the final environment after executing the cinnamon $\Sigma$ with initial environment $\varepsilon$, provided $\Sigma$ halts. If $\Sigma$ does not halt when started in initial environment $\varepsilon$, then $[[\Sigma]] (\varepsilon)$ is undefined. Instead of saying 'Executing the cinnamon $\Sigma$' one can equivalently say 'The cinnamon $\Sigma$ computes' the function $[[\Sigma]]$. This function is the denotational semantics of $\Sigma$. Above, we use emphatic brackets [[ ]] following the tradition in studies on denotational semantics.

The following more formal definition of the execution of each of the allowed primitives follows elements from [13] for the case of while-programs. For $\varepsilon \in Env$, $x \in Varss$, and $a \in N$, let $\varepsilon[x \leftarrow a]$ denote the environment that is identical to $\varepsilon$ except for the value of $x$, which is $a$. Formally,

$$\varepsilon[x \leftarrow a](y) = \varepsilon(y) \text{ if } y \neq x, \text{ and } \varepsilon[x \leftarrow a](x) = a.$$

The execution of the action primitive has the following semantics:

$$[[clear(x)]](\varepsilon) = \begin{cases} \varepsilon[x \leftarrow 0], \text{ if } FORW = 1 \\ \varepsilon, \text{ if } FORW = 0 \end{cases}$$

$$[[copy(x,y)]](\varepsilon) = \begin{cases} \varepsilon[y \leftarrow \varepsilon(x)], \text{ if } FORW = 1 \\ \varepsilon, \text{ if } FORW = 0 \end{cases}$$

$$[[inc(x)]](\varepsilon) = \begin{cases} \varepsilon[x \leftarrow \varepsilon(x)+1], \text{ if } FORW = 1 \\ \varepsilon[x \leftarrow \varepsilon(x)-1], \text{ if } FORW = 0 \end{cases}$$

The execution of the elementary test primitive $nonEq(x,y)$ changes the value of the system variable FAILURE only. Formally, its semantics is as follows:

$$[[noneq(x,y)]](\varepsilon) = \begin{cases} \varepsilon, \text{ if } \varepsilon(x) \neq \varepsilon(y) \ \& \ FORW = 1 \\ \varepsilon[FAILURE \leftarrow 1], \text{ if } \varepsilon(x) = \varepsilon(y) \ \& \ FORW = 1 \\ \varepsilon, \text{ if } FORW = 0 \end{cases}$$

As we discussed, the interpretation of a cinnamon $\Sigma$ as a partial function $[[\Sigma]] : Env \rightarrow Env$. At the same time, for any given natural number j, we might want to consider a cinnamon $\Sigma$ as an agent computing a j-ary partial function $f_\Sigma: N^j \rightarrow N$. If we want to emphasize the arity of the function, we will write $f_\Sigma^{(j)}: N^j \rightarrow N$.

Let a cinnamon $\Sigma$ and a natural number j be given. Let n be the number of variables of the cinnamon $\Sigma$. A partial function $f_\Sigma: N^j \rightarrow N$ defined in the following way, is called **the partial j-ary function computed by $\Sigma$**:

a) Suppose j < n. Let $<a_1, a_2, ..., a_j> \in N^j$. Let $\varepsilon \in Env$ be an environment such that $\varepsilon(x_i) = a_i$ for any i, $1 \leq i \leq j$, $\varepsilon(x_0) = 0$, and $\varepsilon(x_i) = 0$ for any $i > j$. Then $f_\Sigma(a_1, a_2, ..., a_j)$ is defined iff $[[\Sigma]](\varepsilon)(x_0)$ is defined, and if both are defined then

$$f_\Sigma(a_1, a_2, ..., a_j) = [[\Sigma]](\varepsilon)(x_0).$$

b) Suppose $j \geq n$. Let $\varepsilon \in Env$ be an environment such that $\varepsilon(x_i) = a_i$ for any $i \leq n-1$, and $\varepsilon(x_0) = 0$. Then $f_\Sigma(a_1, a_2, ..., a_j)$ is defined iff $[[\Sigma]](\varepsilon)(x_0)$ is defined, and if both are defined then

$$f_\Sigma(a_1, a_2, ..., a_j) = [[\Sigma]](\varepsilon)(x_0).$$

In other words, if $j < n$ where n is the number of variables of $\Sigma$, and $\varepsilon(x_1)$, $\varepsilon(x_2)$, .... $\varepsilon(x_j)$ are the values of the j variables $x_1, ..., x_j$ of $\Sigma$ in the starting environment while the values of the other variables are 0, then $f(\varepsilon(x_1), \varepsilon(x_2), .... \varepsilon(x_j))$ takes the value $[[\Sigma]](\varepsilon(x_0))$ of the variable $x_0$ in the environment after the termination of the computation of $\Sigma$ if it terminates. If the cinnamon has less variables than the parameters of the function then we simply use the first n variables and ignore the remaining ones.

Clearly, a given cinnamon $\Sigma$ computes a nullary (constant) function $f_\Sigma^{(0)}: \{0\} \rightarrow N$, a unary function $f_\Sigma^{(1)}: N \rightarrow N$, a binary function $f_\Sigma^{(2)}: N^2 \rightarrow N$, and so on up to $N^{(n-1)} \rightarrow N$, and further to infinity.

As an example, let us consider the cinnamon from Fig. 3, and let $Var = \{z,x,y,...\}$. Let us first consider the function $f^{(2)}: N^2 \rightarrow N$. It is easy to check that $f^{(2)}(7,3) = 4$, $f^{(2)}(7,7) = 0$, $f^{(2)}(7,0) = 7$, and $f^{(2)}(3,7) = 0$. $f^{(1)}(7) = f^{(2)}(7,0) = 7$. $f^{(0)}() = f^{(2)}(0,0) = 0$. $f^{(3)}(7,3,8) = f^{(2)}(7,3) = 4$. $f^{(4)}(7,3,8,2) = f^{(2)}(7,3) = 4$.

## 4. TURING COMPLETENESS

In this section, we show the Turing-completeness of cinnamon programming. Of course, the power of cinnamon programming cannot be seen and appreciated when computing simple deterministic functions but when dealing with nondeterministic problems. Actually, for deterministic algorithms, cinnamon programming is a departure from the idea of structured programming. However, as cinnamon programming is another computation model, it is still worthy to show its equivalence in computational power to the other, well-established computation models which are all equivalent in power according to the famous Church-Turing thesis.

Computing models are typically considered as transducers (solvers of function problems) or acceptors (solvers of decision problems). A computation model is said to be Turing-complete if its computational power is equivalent to that of the Turing machine, and consequently to all other equivalent models. The computation model which is most similar to cinnamons, is the while-programs. Therefore, we discuss below the equivalence between cinnamons and while-programs. The result is formulated in the next theorem.

> Theorem:
> {i) For every partial function $f: N^j \rightarrow N$ computable by a while-program $P$, there is a cinnamon $\Sigma$ such that $f_\Sigma^{(j)} = f$.
> {ii) For every partial function $f_\Sigma^{(j)}: N^j \rightarrow N$ computable by a cinnamon $\Sigma$, there is a while-program $P$ that computes $f_\Sigma^{(j)}$.

As usual for this type of results, for the proof of part (i) we need to simulate the while-program $P$ by a suitable cinnamon, and to prove part (ii) we need to simulate the cinnamon $\Sigma$ by a suitable while-program.

Slightly different (but equivalent) definitions of the concept of a while-program can be found in the literature. We will use here the definition given in [13]. This definition includes the simple assignments $x := 0$, $x := y+1$ and $x := y$, and four statement constructs: sequential composition $\{p;q\}$, conditional *if x<y then p else q,* for-loop *for y do p,* and while-loop *while x<y do p.* Programs built inductively from these constructs are called while-programs. The function computed by a while-program is defined similarly to that of a cinnamon.

The simulation of all the constructs from the definition of a while-program by cinnamons is shown below:

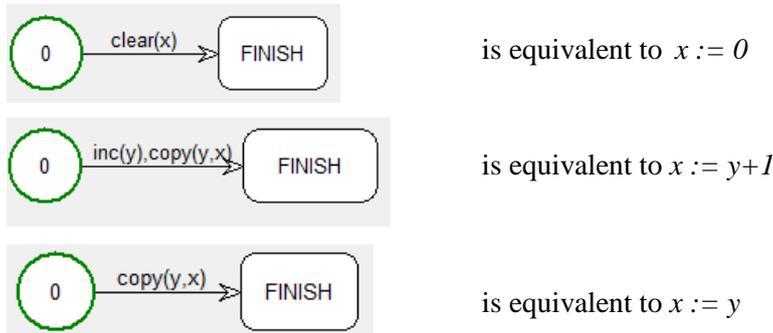

Sequential composition *{P, Q}* from a while program can be simulated in the corresponding cinnamon by simulating separately $P$ and $Q$, and then combining them by unifying the *FINISH* node of the part simulating $P$ with the initial node of the part simulating $Q$. The simulation of the conditional statement *if x < y then P else Q* is illustrated below:

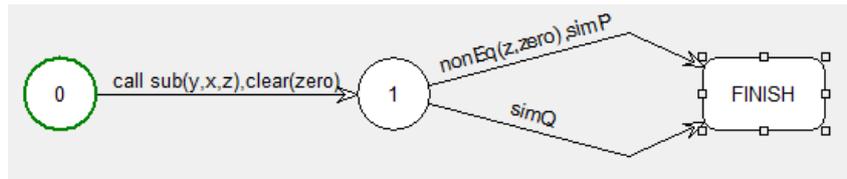

where *SimP* is the cinnamon fragment simulating *P*, and *simQ* is the simulation of *Q*. A for-loop *for y do P* can be simulated by:

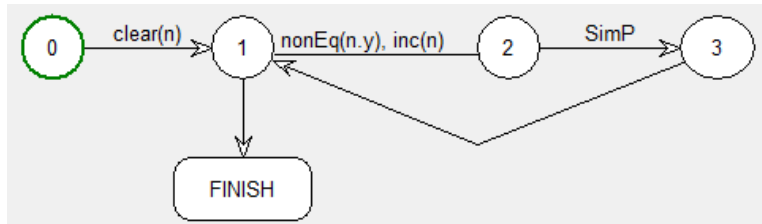

The simulation of *while x<y do P* is illustrated by:

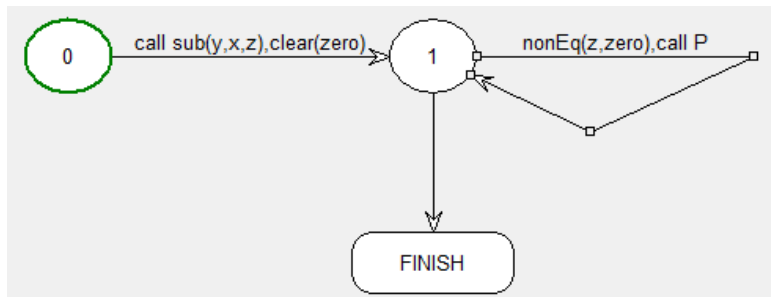

where subnet *P* simulates segment *P* of the while-program.

We will not provide a formal proof of Part (ii) of the Theorem. We can notice, however, that the cinnamon interpreter presented earlier is an algorithm, and as such can be coded in a programming language or by a while-program.

## 5. FULL CNP VS. CINNAMON PROGRAMMING

The cinnamon, or core CNP, is a minimal theoretical computation model underlying the 'real' Control Network Programming. We include below a short list of some extensions and extra features supported by the real CNP.

- Subnets can have local variables.
- In a subnet invocation the programmer can indicate which state of the subnet will be considered as its initial state for the particular subnet call.
- The label of an arrow is a (possibly empty) sequence of primitives and/or subnet calls, not just a single primitive or a single subnet invocation.
- The user defines their own primitives which are arbitrary procedures of the underlying programming language (not only the given minimal set of three elementary action primitives and one test primitive). In addition to forward action, any user-defined primitive can also have backward action. The definitions of the primitives (plus eventually definitions of constants and helping functions) form the operational unit (in *SpiderCNP* it is implemented as a file called *SpiderUnit*). Technically no difference is made between action and test primitives.
- User-defined primitives can be defined in separate units which are linked within the project.

- It should be possible that the source code of these units is written in different programming languages – interoperability.
- Object-oriented programming can be used.
- There exists a system state STOP which is equivalent to a state without outgoing arrows.
- The programmer may define costs of the arrows of the control network.
- User has very powerful control tools – control states and system options – in order to direct and control the CN traversal [24,25]. In particular, that allows simple 'visual' implementation of
  - heuristic algorithms [2,3,26]
  - nondeterministic algorithms [27]
  - randomized algorithms [28]
- The programmer can define the solution scope, that is, how many solutions will be found (if they exist) – a single solution, a fixed in advance number of solutions, all solutions, prompting after each solution if another solution should be sought. For example, the recognizer of arithmetic expressions (Fig. 5) will recognize as legal the expression a+b*c with operation + preceding the operation *, but another solution is possible which corresponds to operation * preceding +.

## 6. CINNAMON PROGRAMMING, NONDETERMINISM, AND OTHER EXTENSIONS

In this section, we address a few ideas for future study.

As discussed in detail in [4] for the case of CNP, the concept of a cinnamon is inherently declarative and nondeterministic in nature. This comes as no surprise at all, as its skeleton is recursive nondeterministic automata – see Section 2 above. At the same time, in Section 3, we defined the cinnamon as a deterministic computation model, and then in Section 4 pointed at its Turing-completeness, that is, its equivalence in computational power to the other deterministic models of computation. It would seem natural to actually define cinnamons as a nondeterministic model (computing a relation rather than a partial function) and relate them to other nondeterministic models such as nondeterministic Turing machines, nondeterministic while-programs [29-33], etc. This direction of study is clearly related to the issue of solution scope. Further, nondeterministic computation is closely related to the definitions of a search problem, and the complexity class NP for search problems – therefore, treatment of cinnamons in this context are possible. This 'nondeterministic' avenue of research, however, as the other suggested topics of this section, lay beyond the scope of the current publication.

As mentioned, the 'real' CNP has control tools that make the 'declarative' implementation of randomized algorithms easy and visually clear. It would be, therefore, reasonable to appropriately extend the concept of a cinnamon and study it as a model of randomized (probabilistic) computation, in particular in relation to existing randomized (probabilistic) computation models and complexity classes [34-36].

Historically, CNP has roots in rule-based systems and fuzzy rule-based systems. Considering fuzzy cinnamons could be another possible direction for study.

**Author**

Dr Kostadin Kratchanov taught at Ruse Univ. (Bulgaria), Technical Univ. of Sofia Plovdiv Branch (Bulgaria), Brunel Univ. (UK), European Univ. of Lefke (Northern Cyprus), Univ. of Bahrain, Grande Prairie Regional Coll. (Canada), Mount Royal Univ. (Canada), and is currently with Yasar Univ. (Turkey). His research interests are in Theoretical Computer Science, Theory of Computation, Discrete Structures, Fuzzy Automata, Applications of Category Theory in Computer Science, Programming Languages and Paradigms, Artificial Intelligence, Analysis and Design of Algorithms, Software Engineering.

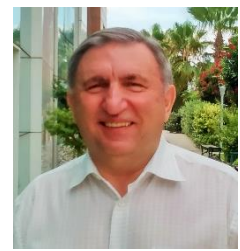